\documentclass[onecolumn,preprintnumbers,superscriptaddress]{revtex4}
\usepackage{epsfig,bm}
\usepackage{ulem}
\usepackage{float,pst-all}
\usepackage{amsmath,color}
\usepackage{gensymb}
\usepackage{subfigure}

\begin{document}

\title{Two-neutron correlations in a Borromean $^{20}{\rm C}+n+n$ system: Sensitivity of unbound subsystems}

\author{Jagjit Singh}
\email{jsingh@rcnp.osaka-u.ac.jp, jsingh@nucl.sci.hokudai.ac.jp}
\affiliation{Nuclear Reaction Data Centre, Faculty of Science, Hokkaido University, Sapporo 060-0810, Japan}
\affiliation{Research Center for Nuclear Physics (RCNP), Osaka University, Ibaraki 567-0047, Japan}
\author{W. Horiuchi}
\email{whoriuchi@nucl.sci.hokudai.ac.jp}
\affiliation{Department of Physics, Hokkaido University, Sapporo 060-0810, Japan}
\author{L. Fortunato}
\email{fortunat@pd.infn.it}
\author{A. Vitturi}
\email{vitturi@pd.infn.it}
\affiliation{Dipartimento di Fisica e Astronomia \textquotedblleft G.Galilei\textquotedblright\\ and INFN-Sezione di Padova, via Marzolo 8, I-35131 Padova, Italy}

\date{\today}

\begin{abstract}
The structure of $^{22}$C plays a vital role in the new physics 
at subshell closure of $N=16$  in the neutron-rich region. 
We study the two-neutron correlations in the ground state of the weakly-bound Borromean
nucleus $^{22}$C sitting at the edge of the neutron-drip line and its sensitivity 
to ${\rm core}$-$n$ potential. For the present study, we employ a three-body (${\rm core}+n+n$) structure model designed
for describing the Borromean system by explicit coupling of unbound continuum states of the subsystem (${\rm core}+n$).
We use a density-independent contact-delta interaction to describe the neutron-neutron interaction and its strength is 
varied to fix the binding energy.
Along with the ground-state properties of $^{22}$C, we investigate its electric-dipole and monopole
responses, discussing the contribution of various configurations.
Our results indicate more configuration mixing as compared to the previous studies in the ground state of $^{22}$C.
However, they strongly depend upon the choice of the $^{20}{\rm C}$-$n$ potential as well as the binding energy of $^{22}$C, 
which call for new precise measurements for the low-lying continuum structure of the binary system ($^{20}{\rm C}+n$) 
and the mass of $^{22}$C. These measurements will be essential to understand the 
Borromean three-body system $^{22}$C with more accuracy.
\end{abstract}
\maketitle
\section{Introduction}
\label{intro}
Nowadays, the structure of the dripline isotopes 
can be studied, due to the novel advancements in the highly sophisticated 
spectrometers to separate out the exotic products of the fragmentation 
reactions between the stable nuclear beams and the production targets. In the exploration 
of the isotopes across the neutron dripline, one of the most 
eye-catching observed phenomena is the neutron halo. A peculiar 
feature of these halo nuclei is the long tail of their matter density distribution. 
In mid of 1980s, at Lawrence Berkeley Laboratory, the first two-neutron halo nucleus 
$^{11}$Li was observed \cite{TANH85}, which has two correlated neutrons in its halo. 
This experiment leads to the conclusion, that weak binding of two neutrons is critical to the formation
of a halo.

Among halo nuclei, the two-neutron halo consists of three pieces, two neutrons 
and a core that bind the system since either of the binary subsystem is unbound. 
Such a three-body quantal system where all three parts must be present for the 
existence of the system is named \textquotedblleft Borromean\textquotedblright \cite{ZHU93}
after the three interlocking rings on the 15th century coat of arms of the Borromeo family in 
northern Italy. 

$^{22}$C ($^{20}{\rm C}+n+n$) is the heaviest observed Borromean nucleus with  
twice as many protons and neutrons as $^{11}$Li has, where neither of the binary system 
$^{20}{\rm C}$-$n$ and $n$-$n$ are bound. This nucleus has gained much attention 
since the early 2000s and is an interesting candidate for checking 
the persistence of $N=16$ magicity, which was established for $^{24}$O \cite{OZA00}. 
Notably, for the persistence of $N=16$ shell closure in $^{22}$C, the two-neutron valence configuration 
would have to be dominated by $s$-wave components, optimal for halo formation.
The experimental evidences for the $s$-wave dominance in the ground state of $^{22}$C which reflects $N=16$ magicity has been observed in the 
two-neutron removal cross section from $^{22}$C and the resulting $^{20}$C 
fragment momentum distribution \cite{KOBA12}. In fact, the estimated (according to the mass evaluation in 2003) 
and the observed (at GANIL by direct time-of-flight method) $2n$-separation energy $(S_{2n})$ 
is very small $:$ $0.42 \pm 0.96 $\,MeV \cite{Audi03} and $0.14 \pm 0.46$\,MeV \cite{GAUD12} with large uncertainties.

However, there is scarce experimental information available before $2010$ 
for $^{22}$C beyond its half-life and $\beta$-delayed neutron multiplicities \cite{YONE03}. 
In $2010$, the first reaction cross section ($\sigma_R$)
measurement of $^{22}$C was performed at RIKEN with a liquid hydrogen $(^1$H) target
 at incident energy of $40$\,MeV/nucleon \cite{TANA10}. 
They observed large $\sigma_R$ for $^{22}$C is $1338 \pm 274$\,mb, showing large 
enhancement with respect to the neighboring lighter carbon isotopes. This results in 
the huge root-mean-square (rms) matter radius of $^{22}$C, i.e., $5.4 \pm 0.9$\,fm with large uncertainties.
In the spirit of attaining high precision, recently the 
interaction cross section ($\sigma_I$) for $^{22}$C was measured at RIKEN with a $^{12}$C target at 
$235$\,MeV/nucleon \cite{TOG16}. The obtained precise value of $\sigma_I$ is $1280 \pm 23$\,mb and 
the resultant rms matter radius is $3.44 \pm 0.08$\,fm, which is much smaller 
than the previous measurement \cite{TANA10}. Very recently, this radius puzzle 
in which two recent interaction cross section measurements
using $^1$H and $^{12}$C targets shows quite different 
radii was discussed \cite{HOR18}. 
On theoretical side, before the first experimental data \cite{TANA10}, the limited theoretical studies 
for the ground-state structure of $^{22}$C and the reaction cross sections 
for $^{22}$C had been reported by Refs.~\cite{HOR06,HOR07,ABU08}. 
These recent experimental measurements \cite{TANA10,TOG16} have triggered 
off several theoretical investigations 
\cite{YAMA12,FORT12,ERSH12,ACHA13,OGA13,INAK14,YLK14,SUZU16,SOUZ16,FORT16,PIN16,HOR18,ERSH18}.

The stability of the three-body ($^{20}{\rm C}+n+n$) system is linked to the continuum 
spectrum of the two-body ($^{20}{\rm C}+n$) subsystem \cite{MIG73}. Thus, the structural 
spectroscopy of $^{21}$C plays a vital role in the understanding of $^{22}$C. 
Experimentally, $^{21}$C is poorly known beyond that it is unbound \cite{MOSB13}. 
The poor information over the structure of $^{21}$C has been used to 
constrain the ${\rm core}+n$ potential which directly leads to uncertainties in the three-body 
calculations of $^{22}$C because the choice of this potential (${\rm core}+n$) 
is linked to the configuration mixing in the ground state of the ${\rm core}+n+n$ system.

Although several three-body calculations have been reported so far 
\cite{HOR06,ERSH12,YLK14,PIN16,ERSH18} with different sets of ${\rm core}$-$n$ potentials, 
most of these calculations showed the $s$-wave dominance in the 
ground state of $^{22}$C, with negligible contribution of 
other higher $l$-wave components \cite{HOR06,ERSH12,YLK14}. But recently the mixing of 
the ground state with $d_{3/2}$ has been 
reported in \cite{PIN16,ERSH18}. Most of these theoretical models explain the ground-state 
structure fairly well to study dynamics of nuclear reactions but the uncertainties over the choice of 
${\rm core}+n$ potentials have not been discussed well so far.

In this paper, we investigate the sensitivity of choice of a ${\rm core}+n$ potential 
with the configuration mixing in the ground state of $^{22}$C, which have not been completely explored. 
We note that the choice of ${\rm core}+n$ potentials in our study 
is different from previous studies \cite{HOR06,PIN16} with the motivation to discuss the effect of 
$l>0$ resonance on the configuration mixing in the ground state of $^{22}$C and also to explore the role of the $l>0$ resonances of the subsystem. 
For this study, we use a three-body (${\rm core}+n+n$) 
structure model for the ground and continuum states of the Borromean nuclei \cite{FORTU14,JS16}. 
We present the ground-state ($J^\pi=0^+$) properties of $^{22}$C and transitions to the continuum 
($J^\pi=0^+$ and $1^-$) that 
might be of help in disentangling the two-neutron correlations in $^{22}$C.
We compare our findings with the experimental and the theoretical works that have been done in 
the recent past on this system.

The paper is organized as follows. 
Section~\ref{MF} briefly describes the formulation of our three-body structure model. 
In Sec.~\ref{21C} we analyze the subsystem $^{21}$C and fix the four different sets for 
${\rm core}+n$ potential, consistent with available scarce experimental and theoretical predictions.
Section~\ref{22C} presents our results for the three-body system, $^{20}{\rm C}+n+n$. 
In Sec.~\ref{MS}, we describe our model space for the three-body system. 
In Sec.~\ref{GSP}, we report our main results of the configuration mixing in the 
ground state of $^{22}$C with different binding energies along with other ground-state properties.
Sections~\ref{DIP} and ~\ref{MON} present the results on electric-dipole and monopole responses of the system, respectively. 
Finally, conclusions are made in Sec.~\ref{CON}. In Appendix we discuss the issue of convergence of our
results with respect to variations in the model parameters.

\section{Model Formulation}

\label{MF}
Here we employ a three-body (${\rm core}+n+n$) structure model 
to study the weakly-bound ground and low-lying continuum states of Borromean systems sitting 
at the edge of neutron dripline \cite{FORTU14,JS16,JS16b,JS16c}. 
We start from the solution of the unbound subsystem (${\rm core}+n$) and the two-particle basis is 
constructed by explicit coupling of the two single-particle continuum wave functions.
Initially, it was tested for studying the structure of $^6$He and has been 
successful in explaining the ground-state properties and the electric-dipole and quadrupole responses \cite{FORTU14,JS16}. 
To confirm the validity of this approach to a heavier Borromean system is another purpose of this paper.

The three-body wave function for the ${\rm core (^{20}C)}+n+n$ system is specified by the Hamiltonian
\begin{equation}
H=-\frac{\hbar^2}{2\mu}\sum_{i=1}^{2}\nabla_i^2+\sum_{i=1}^{2}V_{{\rm core}+n}(\vec r_i)+V_{12}(\vec r_1,\vec r_2)
\end{equation}
where $\mu=A_cm_N/(A_c+1)$ is the reduced mass, and $m_N$ and $A_c=20$ are the nucleon mass and mass 
number of the core nucleus, respectively. The recoil term is neglected in the present study, as $A_c=20$ is large enough to ignore it.
$V_{{\rm core}+n}$ is the core-$n$ potential and $V_{12}$ is $n$-$n$ potential.
The neutron single-particle unbound $s$-, $p$-, $d$- and $f$-wave continuum states of the subsystem ($^{21}$C) 
are calculated in a simple shell model picture for different continuum energy $E_{C}$
by using the Dirac delta normalization and are checked with a more refined phase-shift analysis. 
Each single-particle continuum wave function of $^{21}$C is given by
\begin{equation}
\phi_{\ell j m}(\vec{r},E_C)=R_{\ell j}(r,E_C)[Y_{\ell}(\Omega)\times \chi_{1/2}]^{(j)}_m.\label{spwfn} 
\end{equation}
In the present calculations, we use the mid-point method \cite{AUST87, PIYA99} to discretize
the continuum, which consists of taking the scattering state defined as
\begin{equation}
\tilde{\phi}_i(\vec{r})=\sqrt{\Delta{E}}\;\phi_{\ell jm}(\vec{r},\bar{E}_i) , \quad E_i>0
\end{equation}
for a discrete set of the scattering energies, where $\bar{E}_i=(E_i+E_{i-1})/2$, with $\Delta{E}$ as a common energy 
interval or bin width. In the mid-point method, continuum channels are represented by the channel 
at a midpoint of the bin. The resulting set of wave functions 
$\tilde{\phi}_{ik}(\vec{r})$ satisfies the following orthogonality condition
\begin{equation}
\int\tilde{\phi}_i(\vec{r})\tilde{\phi}_k(\vec{r})d\vec{r}=\Delta{E}\,\delta_{ik}\delta(\bar{E}_i-\bar{E}_k),
\end{equation}  
that depends on the $\Delta{E}$.
The convergence of the results will be checked with the continuum energy cut $E_{\rm\,cut}$ and $\Delta{E}$. 
These ${\rm core}+n$ continuum wave functions are used to construct the 
two-particle $^{22}$C states by proper angular momentum
couplings and taking contribution from different configurations.
The combined tensor product of these two continuum states is given by
\begin{equation}
\psi_{JM}(\vec{r}_1,\vec{r}_2)=[\phi_{\ell_1 j_1 m_1}(\vec r_1,{E_C}_1) \times \phi_{\ell_2 j_2 m_2}(\vec r_2, {E_C}_2)]^{(J)}_M.
\label{tpwfn}
\end{equation}
We use a density-independent (DI) contact-delta pairing interaction for simplicity, 
and its strength is the parameter which will be fixed to reproduce the ground-state energy. 
For $S=0$ the explicit expression for $V_{12}$ is given by
\begin{equation}
V_{12}=g\delta\left(\vec r_1-\vec r_2\right),
\end{equation}
where $g$ is the actual strength of the pairing interaction, is obtained by scaling 
the coefficient of the contact-delta pairing, $G$, 
with the $\Delta E$ \cite{FORTU14}, i.e.,
\begin{equation}
g=\frac{G}{(\Delta E)^2}.
\label{DE}
\end{equation}
We ignore the $S=1$ pairing for the present study. 
The full computational procedure of our approach is described in detail in \cite{JS16b,JS16c}.
In order to check the consistency in the configuration mixing of the ground state of $^{22}$C, 
we have also used a density-dependent (DD) contact-delta pairing interaction \cite{ESBN97}, given by
\begin{equation}
V_{12}=\delta(\vec{r}_1-\vec{r}_2)
\left(v_0+\frac{v_\rho}{1+\exp[(r_1-R_\rho)/a_\rho]}\right).
\label{vnn}
\end{equation}
The first term in Eq.~(\ref{vnn}) with $v_0$ simulates the free $n$-$n$ interaction, which is 
characterized by its strength and the second term in Eq.~(\ref{vnn}) 
represents density-dependent part of the interaction. These strengths $v_0$ and $v_\rho$ are 
also scaled with the $\Delta E$ by following Eq.~(\ref{DE}).
\section{Two-body unbound subsystem ($^{20}{\rm C}+n$)} \label{21C}
The scrutiny of the $^{20}{\rm C}+n$ subsystem ($^{21}$C) 
is imperative in studying $^{22}$C as 
a typical nucleus of Borromean family as $^{20}{\rm C}+n+n$. 
The interaction of the $^{20}$C core 
with the valence neutron ($n$) plays a crucial role in the binding mechanism of $^{22}$C.
The unbound nucleus $^{21}$C can be described as an inert $^{20}$C core with an unbound 
neutron moving in $s$-, $p$-, $d$- or $f$-wave continuum states in a simple independent-particle shell model picture. 
The subshell closure of the neutron number $14$ is assumed for the core configuration 
given by $(0s_{1/2})^2(0p_{3/2})^4(0p_{1/2})^2(0d_{5/2})^6$. 
The seven valence neutron continuum orbits, i.e., $s_{1/2}$, $d_{3/2}$, $f_{7/2}$, $p_{3/2}$, $f_{5/2}$, $p_{1/2}$ and $d_{5/2}$
are considered in the present calculations for $^{21}$C.
The primary issue over the choice of a ${\rm core}+n$ potential is the scarce experimental information about the neutron-core system.
The only available experimental study using the single-proton removal reaction 
from $^{22}$N at $68$\,MeV/nucleon reported the limit to the scattering 
length $\lvert a_0\lvert < 2.8$\,fm \cite{MOSB13}. In addition, no 
evidence was found for a low lying state.
Due to the low accuracy of the experimental data \cite{MOSB13} at low energies, the possibility of resonance states
can not be ruled out. The possibility of $d_{3/2}$ resonance has been discussed in \cite{OGA13,PIN16}. 
We construct the ${\rm core}+n$ potential in the view of fixing the available data for the 
scattering length \cite{MOSB13} and 
the energy of the $d_{3/2}$ resonance in Ref.~\cite{PIN16} and references therein.

Assuming the $^{20}$C core to be inert in the ground state, we take the following ${\rm core}+n$ potential
\begin{equation}
V_{{\rm core}+n} = \left(V_0^l+V_{ls}\vec{l}\cdot\vec{s}\frac{1}{r}\frac{d}{dr}\right) \frac{1}{1+{\rm\,exp}(\frac{r-R_c}{a})},
\label{vsp}
\end{equation}
where $R_c=r_0A_c^{\frac{1}{3}}$ ($A_c=20$ for $^{20}$C) with $r_0=1.25$\,fm and $a=0.65$\,fm. 

\begin{table}[h!]
\centering
\caption{Parameter sets of the ${\rm core}$-$n$ potential for $l=0, 1, 2, 3$ states of a $^{20}$C$+n$ system. 
The scattering length $a_0$ in fm and possible resonances with the resonance energy $E_{R}$ and the decay width $\Gamma$ in 
MeV are also tabulated. The parameters, $a$ and $r_0$ are $0.65$ and $1.25$\,fm for all the four sets. See text for details.}
\scalebox{1.1}{
\begin{tabular}{c|ccccccc}
\hline\hline
{Set}&\multicolumn{1}{c}{$l$} & \multicolumn{1}{c}{ $V_0^l$ (MeV)} & \multicolumn{1}{c}{$V_{ls}$ (MeV)}&\multicolumn{1}{c}{{$l_j$}}&\multicolumn{1}{c}{$a_0$}&\multicolumn{1}{c}{$E_{R}$}&\multicolumn{1}{c}{ $\Gamma$} \\ \hline
${1}$&\multicolumn{1}{c}{$0$}&\multicolumn{1}{c}{$-33.54$}& \multicolumn{1}{c}{$0.00$}&\multicolumn{1}{c}{$s_{1/2}$}&\multicolumn{1}{c}{$-2.8$}&  \multicolumn{1}{c}{...}                  & \multicolumn{1}{c}{...}       \\
&\multicolumn{1}{c}{$1,2,3$}               & \multicolumn{1}{c}{$-43.24$}     & \multicolumn{1}{c}{$25.63$} &\multicolumn{1}{c}{$d_{3/2}$}& \multicolumn{1}{c}{...} &\multicolumn{1}{c}{1.51}   &\multicolumn{1}{c}{0.38}       \\ 
&\multicolumn{1}{c}{} &\multicolumn{1}{c}{} &\multicolumn{1}{c}{} &\multicolumn{1}{c}{$f_{7/2}$} & \multicolumn{1}{c}{...}                  & \multicolumn{1}{c}{7.42}  & \multicolumn{1}{c}{4.05}     \\ \hline
${2}$&\multicolumn{1}{c}{$0$} &  \multicolumn{1}{c}{$-33.54$}     & \multicolumn{1}{c}{$0.00$}      &\multicolumn{1}{c}{$s_{1/2}$}& \multicolumn{1}{c}{$-2.8$} &\multicolumn{1}{c}{...}                   & \multicolumn{1}{c}{...}       \\
&\multicolumn{1}{c}{$1,2,3$} &  \multicolumn{1}{c}{$-45.14$}     & \multicolumn{1}{c}{$25.63$}   & \multicolumn{1}{c}{$d_{3/2}$} & \multicolumn{1}{c}{...}& \multicolumn{1}{c}{$0.83$}      & \multicolumn{1}{c}{$0.09$}              \\ 
&\multicolumn{1}{c}{} &\multicolumn{1}{c}{} &\multicolumn{1}{c}{} &\multicolumn{1}{c}{$f_{7/2}$} & \multicolumn{1}{c}{...}& \multicolumn{1}{c}{$6.67$} & \multicolumn{1}{c}{$2.93$}         \\ \hline
${3}$&\multicolumn{1}{c}{$0$} & \multicolumn{1}{c}{$-33.00$}     & \multicolumn{1}{c}{$0.00$}  & \multicolumn{1}{c}{$s_{1/2}$}& \multicolumn{1}{c}{$-47.6$}&\multicolumn{1}{c}{...}&\multicolumn{1}{c}{...}        \\
&\multicolumn{1}{c}{$2$} &  \multicolumn{1}{c}{$-47.50$}     & \multicolumn{1}{c}{$35.00$} & \multicolumn{1}{c}{$d_{3/2}$} & \multicolumn{1}{c}{...}& \multicolumn{1}{c}{$0.92$}&\multicolumn{1}{c}{$0.09$}        \\ 
&\multicolumn{1}{c}{$1,3$}& \multicolumn{1}{c}{$-42.00$}     & \multicolumn{1}{c}{$35.00$}  &   \multicolumn{1}{c}{$f_{7/2}$}& \multicolumn{1}{c}{...} & \multicolumn{1}{c}{$6.69$}                    & \multicolumn{1}{c}{$3.02$}      \\ \hline
${4}$&\multicolumn{1}{c}{$0$} &  \multicolumn{1}{c}{$-33.54$}     & \multicolumn{1}{c}{$0.00$}      &\multicolumn{1}{c}{$s_{1/2}$}& \multicolumn{1}{c}{$-2.8$} & \multicolumn{1}{c}{...}                 & \multicolumn{1}{c}{...}     \\
&\multicolumn{1}{c}{$1,2,3$} &  \multicolumn{1}{c}{$-47.80$}     & \multicolumn{1}{c}{$35.00$}   & \multicolumn{1}{c}{$d_{3/2}$} & \multicolumn{1}{c}{...}& \multicolumn{1}{c}{$0.80$}                    & \multicolumn{1}{c}{$0.08$}              \\ 
& \multicolumn{1}{c}{}& \multicolumn{1}{c}{}& \multicolumn{1}{c}{}&\multicolumn{1}{c}{$f_{7/2}$} &\multicolumn{1}{c}{...}& \multicolumn{1}{c}{$4.66$}                   & \multicolumn{1}{c}{$1.03$}         \\ \hline\hline
\end{tabular}}
\label{T2}
\end{table}
Considering the sensitivity of the ${\rm core}$-$n$ potential, we examine four 
different potential sets for the present study listed in Table~\ref{T2}. 
Set~$1$ of our present work is basically 
the same as  Set~$\rm\,B$ of \cite{HOR06} and the only difference is that, 
for $s$-wave, $V_0$ is chosen to keep $^{21}$C unbound and the corresponding 
calculated scattering length is $-2.8$\,fm, 
which is consistent with the available experimental information \cite{MOSB13}.

In order to include the $d_{3/2}$ resonance at the position predicted
in \cite{PIN16}, we modify the potential given in \cite{HOR06}, i.e., our Set~$2$, 
by making the potential deeper for $l>0$ waves, we find that the 
narrow $d$-wave resonance is displaced to lower energy at $0.83$ MeV with the decay width 
of $0.09$ MeV. These numbers are consistent with the results presented in \cite{PIN16}. 
In our strategy, we use the same potential for all other $l$-waves ($l>0$). 
In addition to the $d_{3/2}$ resonance, we find a wider $f_{7/2}$ resonance at 
relatively lower energy than Set~$1$, i.e., at $6.67$ MeV with the decay width 
of $2.93$ MeV.
For Sets~$1$ and $2$, the spin-orbit strength is fixed to $V_{ls}=25$ MeV, whereas in 
\cite{PIN16}, the spin-orbit strength is fixed to $V_{ls}=35$ MeV. In order to explore the 
position and the decay width of the $d_{3/2}$ and $f_{7/2}$ resonances, 
our analysis includes two more sets with $V_{ls}=35$ MeV.

Our Set~$3$ is exactly the same as ``set~$3$'' of Ref.~\cite{PIN16}. 
With this potential the scattering length $-47.6$\,fm  and the $d_{3/2}$ resonance energy are reproduced. 
As it can be seen from Set~$3$ that different depths for different $l$-waves are employed. 
They used a less deep potential for the $p$- and $f$-waves than for the $d$-wave.
Basically, Sets~$1$ and $3$ are introduced to compare the consistency of the configuration mixing in the ground state of $^{22}$C 
predicted by the other approaches \cite{HOR06,PIN16} with ours.

With the motivation of using only one-potential depth for $l>0$ waves, we modify the potential given in \cite{PIN16}, 
see Set~$4$ in Table~\ref{T2}. With this modified set, we get the 
$d_{3/2}$ resonance at $0.80$\,MeV, which is consistent with ``set~$1$'' of Ref.~\cite{PIN16}. Interestingly, with this 
set we also find the wider $f_{7/2}$ resonance as with the other 
Sets of Table~\ref{T2} but with this spin-orbit strength it is shifted to the lower energy at $4.66$ MeV with 
the decay width of $1.03$ MeV. The very recent work by Shulgina {\it et al.} \cite{ERSH18} deserves a particular mention, 
as it discussed the sensitivity of the depth of the ${\rm core}$-$n$ 
potential and reported the configuration mixing in the ground state of $^{22}$C only with $V_{ls}=25$\,MeV. 

\section{Three-body system ($^{20}{\rm C}+n+n$)}\label{22C} 
\subsection{Model space}\label{MS} 
As all of these few-body models \cite{HOR06,YLK14,PIN16,ERSH18} described the structure of $^{22}$C to a 
reasonable degree, these approaches normally take as a starting point 
for calculations with a basis set of bound wave functions which damp at large distances. 
In our approach, we calculate the full
continuum single-particle spectrum of $^{21}$C in a straightforward
fashion and use two oscillating continuum wave
functions to construct the two-particle states. 

The relative motion of the neutron 
with respect to the core is an unbound ($E_C>0, k> 0$), 
oscillating wave that must approximate a combination of spherical Bessel functions at large distances from the center.
In the present study, the continuum single-particle wave functions are calculated, 
normalized to a Dirac delta in energy, for the $s$-, $p$-, $d$- and $f$-wave continuum states of $^{21}$C on a radial grid that goes from $0.1$\,--$100.0$\,fm with the 
four different potential sets presented in Table~\ref{T2}.

The three-body model with two non-interacting particles in the above single-particle levels of $^{21}$C produces
different parity states, when two neutrons are placed in seven different unbound orbits, i.e.,
$s_{1/2}$, $d_{3/2}$, $f_{7/2}$, $p_{3/2}$, $f_{5/2}$, $p_{1/2}$ and $d_{5/2}$.
Namely seven configurations $(s_{1/2})^2$, $(p_{1/2})^2$, $(p_{3/2})^2$, $(d_{3/2})^2$,
$(d_{5/2})^2$, $(f_{5/2})^2$ and $(f_{7/2})^2$ couple to $J^\pi=0^+$, 
eight configurations $(s_{1/2}p_{1/2})$, $(s_{1/2}p_{3/2})$, $(p_{1/2}d_{3/2})$, $(p_{3/2}d_{3/2})$, 
$(p_{3/2}d_{5/2})$, $(d_{3/2}f_{5/2})$, $(d_{5/2}f_{5/2})$ and 
$(d_{5/2}f_{7/2})$ couple to $J^\pi=1^-$. 

In the three-body calculations, along with the core-$n$ potential the other important ingredient is the $n$-$n$ interaction.
The single-particle states must necessarily have a positive energy for each type of relative motion angular momentum. 
One must resort to the binding effect of some residual interaction, that 
brings one of the many unbound energy eigenvalues down into
the bound regime following \cite{VITT10,HAG11} for the ground-state $J^\pi=0^+$ case.
An attractive contact-delta pairing interaction is used, 
{$g\delta(\vec r_1 - \vec r_2)$} for simplicity, 
with only one adjustable parameter, $G$, that is related to $g$ with Eq.~(\ref{DE}).

The task of introduction of the residual $n$-$n$ interaction  between the continuum states, 
requires careful numerical implementation because one deals with large data sets. 
The resulting pairing matrix is diagonalized with standard routines and it gives the eigenvalues for the $J=0^+$ case. 
The coefficient of the contact-delta pairing, $G$, is adjusted to reproduce the correct ground-state energy each time. 
The actual pairing interaction $g$ is obtained by correcting with a factor that depends on the 
aforementioned spacing between energy states, i.e., $\Delta{E}$. The biggest adopted basis size gives a fairly dense continuum in 
the region of interest. All the calculations discussed in this paper are performed with $\Delta{E}=0.1$\,MeV and 
$E_{\rm{cut}}=5$\,MeV. For detailed discussion on convergence of these model parameters 
one can refer to the Appendix. 
\subsection{Ground-state properties of $^{22}$C}\label{GSP}
The ground-state wave function obtained from the diagonalization in the 
sufficiently large adopted basis, 
shows a certain degree of collectivity, taking contributions of comparable magnitude from several basis states, 
while in contrast the remaining unbound states usually are made up of a few major components. 
We calculate the wave functions for shallow ($-$0.140\,MeV) 
and deep ($-$0.500\,MeV) binding cases.
The detailed components for each configuration are 
summarized in Table~\ref{CMC22a}, for the shallow case.

\begin{table}[h] 
\centering
\caption{Components of the ground state ($0^+$) of $^{22}$C in full model space, i.e., $s$, $p$, $d$ and $f$-waves 
for the shallow case with the ground-state energy $-0.140$\,MeV.}
\scalebox{1.1}{
\begin{tabular}{c|c|c|c|c}
\hline \hline
\multicolumn{1}{c}{{$(l_j)^2$}}      & \multicolumn{1}{c}{{Set~$1$}} &\multicolumn{1}{c}{{Set~$2$}}&\multicolumn{1}{c}{{Set~$3$}}& \multicolumn{1}{c}{{Set~$4$}} \\
\hline
\multicolumn{1}{c}{$(s_{1/2})^2$}&\multicolumn{1}{c}{$0.923$}&\multicolumn{1}{c}{$0.817$}&\multicolumn{1}{c}{$0.857$}&\multicolumn{1}{c}{$0.553$}\\
\multicolumn{1}{c}{$(p_{1/2})^2$}&\multicolumn{1}{c}{$0.004$}&\multicolumn{1}{c}{$0.004$}&\multicolumn{1}{c}{$0.003$}&\multicolumn{1}{c}{$0.002$}\\
\multicolumn{1}{c}{$(p_{3/2})^2$}&\multicolumn{1}{c}{$0.022$}&\multicolumn{1}{c}{$0.026$}&\multicolumn{1}{c}{$0.021$}&\multicolumn{1}{c}{$0.038$}\\
\multicolumn{1}{c}{$(d_{3/2})^2$}&\multicolumn{1}{c}{$0.045$}&\multicolumn{1}{c}{$0.132$}&\multicolumn{1}{c}{$0.078$}&\multicolumn{1}{c}{$0.083$}\\
\multicolumn{1}{c}{$(d_{5/2})^2$}&\multicolumn{1}{c}{$0.001$}&\multicolumn{1}{c}{$0.001$}&\multicolumn{1}{c}{$0.001$}&\multicolumn{1}{c}{$0.001$}\\
\multicolumn{1}{c}{$(f_{5/2})^2$}&\multicolumn{1}{c}{$0.0003$}&\multicolumn{1}{c}{$0.0002$}&\multicolumn{1}{c}{$0.0002$}&\multicolumn{1}{c}{$0.0001$}\\
\multicolumn{1}{c}{$(f_{7/2})^2$}&\multicolumn{1}{c}{$0.003$}&\multicolumn{1}{c}{$0.017$}&\multicolumn{1}{c}{$0.037$}&\multicolumn{1}{c}{$0.321$}\\
\hline
\multicolumn{1}{c}{Total}&\multicolumn{1}{c}{$1.000$}&\multicolumn{1}{c}{$0.999$}&\multicolumn{1}{c}{$1.000$}&\multicolumn{1}{c}{$1.000$}\\
\hline
\multicolumn{1}{c}{Pairing strength $G$}&\multicolumn{1}{c}{$-7.333$}&\multicolumn{1}{c}{$-6.086$}&\multicolumn{1}{c}{$-5.975$}&\multicolumn{1}{c}{$-3.892$}\\\hline\hline
\end{tabular}}
\label{CMC22a}
\end{table}
 As it can be clearly seen from the Table~\ref{CMC22a}, on moving from Set~$1$ to Set~$4$, the mixing in the 
 ground state with higher $l$-wave components 
 increases in magnitude. Our results for smaller binding energy ($-0.140$\,MeV) shows the $s$-wave dominance for all sets, 
 which is consistent with experimental observation \cite{KOBA12}. 
 In order to explore the dependence of the configuration mixing with the binding energy, 
 we also fix the ground-state energy deeper, i.e., $-0.500$\,MeV. 
 As summarized in Table~\ref{CMC22b}, we find that for Sets~$1$--$3$, the $s$-wave dominance is present, whereas 
 for Set~$4$ an inversion of $s$- and $f$-wave components takes place.  
\begin{table}[h] 
\centering
\caption{Same as Table~\ref{CMC22a} but for the deep case with the ground-state energy $-0.500$\,MeV.}
\scalebox{1.1}{
 \begin{tabular}{c|c|c|c|c}
\hline \hline
\multicolumn{1}{c}{{$(l_j)^2$}}      & \multicolumn{1}{c}{{Set~$1$}} &\multicolumn{1}{c}{{Set~$2$}}&\multicolumn{1}{c}{{Set~$3$}}& \multicolumn{1}{c}{{Set~$4$}} \\
\hline
\multicolumn{1}{c}{$(s_{1/2})^2$}&\multicolumn{1}{c}{$0.856$}&\multicolumn{1}{c}{$0.667$}&\multicolumn{1}{c}{$0.729$}&\multicolumn{1}{c}{$0.358$}\\
\multicolumn{1}{c}{$(p_{1/2})^2$}&\multicolumn{1}{c}{$0.006$}&\multicolumn{1}{c}{$0.006$}&\multicolumn{1}{c}{$0.004$}&\multicolumn{1}{c}{$0.002$}\\
\multicolumn{1}{c}{$(p_{3/2})^2$}&\multicolumn{1}{c}{$0.035$}&\multicolumn{1}{c}{$0.038$}&\multicolumn{1}{c}{$0.032$}&\multicolumn{1}{c}{$0.044$}\\
\multicolumn{1}{c}{$(d_{3/2})^2$}&\multicolumn{1}{c}{$0.092$}&\multicolumn{1}{c}{$0.252$}&\multicolumn{1}{c}{$0.154$}&\multicolumn{1}{c}{$0.118$}\\
\multicolumn{1}{c}{$(d_{5/2})^2$}&\multicolumn{1}{c}{$0.002$}&\multicolumn{1}{c}{$0.002$}&\multicolumn{1}{c}{$0.002$}&\multicolumn{1}{c}{$0.001$}\\
\multicolumn{1}{c}{$(f_{5/2})^2$}&\multicolumn{1}{c}{$0.0004$}&\multicolumn{1}{c}{$0.0003$}&\multicolumn{1}{c}{$0.0003$}&\multicolumn{1}{c}{$0.0001$}\\
\multicolumn{1}{c}{$(f_{7/2})^2$}&\multicolumn{1}{c}{$0.006$}&\multicolumn{1}{c}{$0.034$}&\multicolumn{1}{c}{$0.076$}&\multicolumn{1}{c}{0.475}\\
\hline
\multicolumn{1}{c}{Total}&\multicolumn{1}{c}{$1.000$}&\multicolumn{1}{c}{$0.999$}&\multicolumn{1}{c}{$1.000$}&\multicolumn{1}{c}{$1.000$}\\
\hline
\multicolumn{1}{c}{Pairing strength $G$}&\multicolumn{1}{c}{$-8.903$}&\multicolumn{1}{c}{$-7.079$}&\multicolumn{1}{c}{$-6.925$}&\multicolumn{1}{c}{$-4.283$}\\\hline\hline
\end{tabular}}
\label{CMC22b}
\end{table}
The present calculations of the configuration mixing percentage is consistent with the previous 
calculations of \cite{HOR06} for Set~$1$ and of \cite{PIN16} for Set~$3$, 
although these \cite{HOR06,PIN16} and all the other three-body calculations used more sophisticated two-neutron 
interactions. Here we want to stress that, the comparison of our Set~$4$ results cannot be made with the results of \cite{PIN16}, 
where the authors use different potential strengths for different $l$-waves, whereas in our case the potential strengths for all $l>0$ waves are the same. 
That's why we have $f$-wave dominance in contrast to their findings \cite{PIN16}. 
We also use the DD contact-delta pairing interaction \cite{ESBN97} to check the consistency 
with the DI contact-delta pairing interaction. 
In the DD contact-delta pairing interaction of Eq.~(\ref{vnn}), the strength of the DI part is given
as \cite{ESBN97}
\begin{equation} 
v_0=2\pi^2\frac{\hbar^2}{m_N}\,\frac{2a_{nn}}{\pi-2k_ca_{nn}}, 
\end{equation}
where $a_{nn}$ is the scattering length for the free neutron-neutron scattering
and $k_c$ is related to the cutoff energy, ${{e}_{c}}$, as 
$k_c=\sqrt{\dfrac{m_Ne_{c}}{\hbar^2}}$.
We use $a_{nn}=−15$\,fm and $e_{c}=30$\,MeV \cite{ESBN97}, which leads to $v_0=857.2$\,MeV\,fm$^3$. 
For the parameters of the DI part in Eq.~(\ref{vnn}), we
determine them so as to fix the ground-state energy of
$^{22}$C, $E=-0.140$ MeV \cite{GAUD12}. The values of the parameters that we
employ are $R_\rho=1.25\times A_c^{\frac{1}{3}}$ ($A_c=20$), $a=0.65$\,fm. and $v_\rho=581.55$\,MeV\,fm$^3$ . 
We found that configuration mixing in the ground state of $^{22}$C 
does not change much with the choice of interaction. 
We present the numbers for our potential Set~$3$ in Table~\ref{DIDD}, 
and the same behavior is observed for the other sets, too.

\begin{table}[h]
\centering
\caption{Comparison of the components of the ground state ($0^+$) of $^{22}$C for the DI and DD pairing interactions.
Calculations are done in the full model space 
for the shallow case with the ground-state energy $-0.140$\,MeV.}
\centering
\scalebox{1.1}{
\begin{tabular}{c|c|c|c}
\hline\hline
 \multicolumn{1}{c}{$\Delta E$}& \multicolumn{1}{c}{$(l_j)^2$}        & \multicolumn{1}{c}{$DI$} &\multicolumn{1}{c}{$DD$} \\\hline
\multicolumn{1}{c}{}&\multicolumn{1}{c}{$(s_{1/2})^2$}&\multicolumn{1}{c}{$0.857$}&\multicolumn{1}{c}{$0.819$}  ~\\ 
\multicolumn{1}{c}{}&\multicolumn{1}{c}{$(p_{1/2})^2$}&\multicolumn{1}{c}{$0.003$}&\multicolumn{1}{c}{$0.006$}  ~\\ 
\multicolumn{1}{c}{}&\multicolumn{1}{c}{$(p_{3/2})^2$}&\multicolumn{1}{c}{$0.021$}&\multicolumn{1}{c}{$0.035$}   ~\\ 
\multicolumn{1}{c}{0.1}&\multicolumn{1}{c}{$(d_{3/2})^2$}&\multicolumn{1}{c}{$0.078$}&\multicolumn{1}{c}{$0.084$}        ~\\ 
\multicolumn{1}{c}{}&\multicolumn{1}{c}{$(d_{5/2})^2$}&\multicolumn{1}{c}{$0.001$}&\multicolumn{1}{c}{$0.003$}  ~\\ 
\multicolumn{1}{c}{}&\multicolumn{1}{c}{$(f_{5/2})^2$}&\multicolumn{1}{c}{$0.001$}&\multicolumn{1}{c}{$0.001$} ~\\ 
\multicolumn{1}{c}{}&\multicolumn{1}{c}{$(f_{7/2})^2$}&\multicolumn{1}{c}{$0.037$}&\multicolumn{1}{c}{$0.049$}~\\\hline
\multicolumn{1}{c}{Total}&\multicolumn{1}{c}{}&\multicolumn{1}{c}{$0.999$}&\multicolumn{1}{c}{$1.000$}~\\\hline\hline
\end{tabular}}
\label{DIDD}
\end{table}
 
We find that the percentage of different configuration mixing in the ground state of $^{22}$C, 
are consistent with \cite{HOR06} for Set~$1$ and with \cite{PIN16}
for Set~$3$. 
The calculated ground-state properties are summarized for all the Sets in Tables~\ref{gsp} and \ref{gsp2}  in 
comparison with the experimental \cite{TOG16} and theoretical values \cite{HOR18}.
\begin{table}[h]
\centering
\caption{Radial extension of the ground state of $^{22}$C in units of fm, in the full model space, 
for the shallow case with the ground-state energy $-0.140$\,MeV.}
\scalebox{1.1}{
\begin{tabular}{c|cccccc}
\hline \hline           
\multicolumn{1}{c}{}                     & \multicolumn{1}{c}{Set~$1$} & \multicolumn{1}{c}{Set~$2$}  & \multicolumn{1}{c}{Set~$3$}& \multicolumn{1}{c}{Set~$4$}   &  \multicolumn{1}{c}{Expt. \cite{TOG16}} &  \multicolumn{1}{c}{Ref.~\cite{HOR18}}  \\ \hline
\multicolumn{1}{c}{$R_{\rm m}$}          & \multicolumn{1}{c}{3.35} & \multicolumn{1}{c}{3.32}   &  \multicolumn{1}{c}{3.31} & \multicolumn{1}{c}{3.51}  &    $3.44\pm 0.08$ & $3.38\pm 0.10$ \\
\multicolumn{1}{c}{$\sqrt{\langle r_{NN}^2\rangle}$}   & \multicolumn{1}{c}{8.42} & \multicolumn{1}{c}{8.18}   & \multicolumn{1}{c}{8.06}  & \multicolumn{1}{c}{9.87}  &    &  \\
\multicolumn{1}{c}{$\sqrt{\langle r_{c-2N}^2\rangle}$} & \multicolumn{1}{c}{4.21} & \multicolumn{1}{c}{4.10}   & \multicolumn{1}{c}{4.03}  & \multicolumn{1}{c}{4.93}  &    &   \\ \hline  \hline                         
\end{tabular}}
\label{gsp}
\end{table}

\begin{figure}[h!]
\begin{center}
   \subfigure[Set~$2$] {\epsfig{file=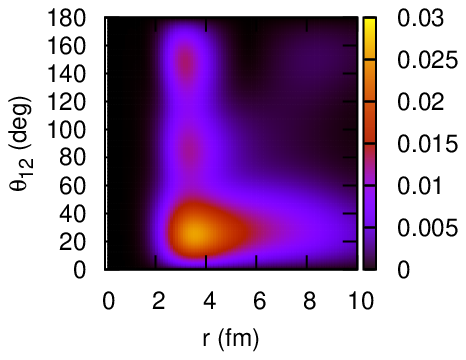, scale=1.2}}
    \subfigure[Set~$4$] {\epsfig{file=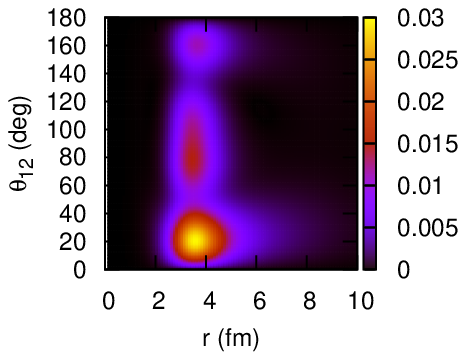, scale=1.2}}
    \caption{
      (Color online) Two-particle density for the ground state of $^{22}$C as a function 
      of $r_1=r_2=r$ and the opening angle between the valence neutrons $\theta_{12}$ 
      for the shallow case with the ground-state energy, $-0.140$\,MeV.
      Set~$2$ and $4$ are employed.
      }
    \label{DenS}
  \end{center}
      \end{figure}
The matter radius is calculated by
\begin{align}
\langle R_{m}^2\rangle=&\dfrac{A_{\rm core}}{A_{\rm core}+2}\langle r_{\rm core}^{2}\rangle+\frac{1}{A_{\rm core}+2}\nonumber\\
&\left(\frac{1}{2}\langle r_{NN}^2\rangle+\frac{2A_{\rm core}}{A_{\rm core}+2}\langle r^2_{c-NN}\rangle\right),\label{mr3}
\end{align}
where 
$\sqrt{\langle r_{\rm core}^2\rangle}=2.99\,{\rm fm}$ \cite{HOR06} and this value is consistent with the evaluated value, i.e., $2.98\pm0.05\,{\rm fm}$ \cite{OZA01}.
The mean-square distance between the valence neutrons and 
that of their center of mass with respect to the core are respectively given by
\begin{align}
\langle r_{NN}^2\rangle &= \langle \Psi_{00}(\vec{r}_1,\vec{r}_2) |(\vec{r}_1-\vec{r}_2)^2| \Psi_{00}(\vec{r}_1,\vec{r}_2) \rangle,\nonumber\\
\langle r_{c-NN}^2\rangle &= \langle \Psi_{00}(\vec{r}_1,\vec{r}_2) |(\vec{r}_1+\vec{r}_2)^2/4| \Psi_{00}(\vec{r}_1,\vec{r}_2) \rangle,
\end{align}
where $\Psi_{00}(\vec{r}_1,\vec{r}_2)$ is the ground-state wave function.
\begin{table}[h]
\centering
\caption{Same as Table~\ref{gsp} but for the deep case with the ground-state energy $-0.500$\,MeV.}
\scalebox{1.1}{
\begin{tabular}{c|cccccc}
\hline \hline           
\multicolumn{1}{c}{}                     & \multicolumn{1}{c}{Set~$1$} & \multicolumn{1}{c}{Set~$2$}  & \multicolumn{1}{c}{Set~$3$}& \multicolumn{1}{c}{Set~$4$}   &  \multicolumn{1}{c}{Expt. \cite{TOG16}} &  \multicolumn{1}{c}{Ref.~\cite{HOR18}}  \\ \hline
\multicolumn{1}{c}{$R_{\rm m}$}          & \multicolumn{1}{c}{3.18} & \multicolumn{1}{c}{3.14}   &  \multicolumn{1}{c}{3.17} & \multicolumn{1}{c}{3.32}  &    $3.44\pm 0.08$ & $3.38\pm 0.10$ \\
\multicolumn{1}{c}{$\sqrt{\langle r_{NN}^2\rangle}$}   & \multicolumn{1}{c}{6.78} & \multicolumn{1}{c}{6.30}   & \multicolumn{1}{c}{6.61}  & \multicolumn{1}{c}{8.13}  &    &  \\
\multicolumn{1}{c}{$\sqrt{\langle r_{c-2N}^2\rangle}$} & \multicolumn{1}{c}{3.39} & \multicolumn{1}{c}{3.15}   & \multicolumn{1}{c}{3.30}  & \multicolumn{1}{c}{4.06}  &    &   \\ \hline  \hline                         
\end{tabular}}
\label{gsp2}
\end{table}

\begin{figure}[h!]
\begin{center}
   \subfigure[Set~$2$] {\epsfig{file=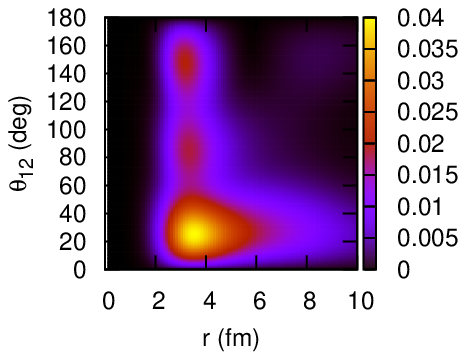, scale=1.2}}
    \subfigure[Set~$4$] {\epsfig{file=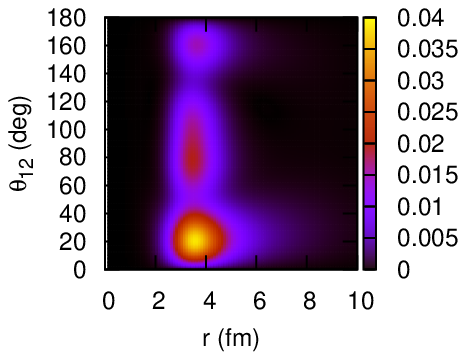, scale=1.2}}
    \caption{
      (Color online) Same as Fig.~(\ref{DenS}) but  
      for the deep case with the ground-state energy, $-0.500$\,MeV.
      Set~$2$ and $4$ are employed.
      }
    \label{DenD}
  \end{center}
\end{figure}

In Tables~\ref{gsp} and \ref{gsp2}, while the matter radius is fairly consistent with the experimental value \cite{TOG16} 
and recent theoretical evaluation \cite{HOR18} for the shallow case, for the deep case it is slightly underestimated for all sets, 
but our values for deep case are consistent with the mean field calculations \cite{INAK14,SUN18}. 
There are slight differences for $\sqrt{\langle r_{NN}^2\rangle}$ and $\sqrt{\langle r_{c-2N}^2\rangle}$ in comparison to previous studies \cite{HOR06,ERSH18} that 
can be ascribed to the choice of different pairing interactions. 
However, these properties are not sensitive to the configuration mixing as reported in Tables~\ref{CMC22a} and \ref{CMC22b}, 
due to difference in magnitude of the pairing strength. 
In the shallow case, the calculated matter radii are closer to the recent theoretical evaluation \cite{HOR18} and the 
the experimental observation \cite{TOG16} than those of the deep case. 
So, our results favor the shallow binding energy and they call for the accurate experimental data for 
$^{22}$C binding energy as there are large uncertainties reported in the previous measurement \cite{GAUD12}. 

The two particle density of $^{22}$C as a function of two radial coordinates, $r_1$ and $r_2$, 
for valence neutrons, and the angle between them, $\theta_{12}$ in the LS-coupling scheme 
is calculated by following Refs.~\cite{JS16,HAG11}.

\begin{figure}[h!]
\centering
{\includegraphics[width=0.6\textwidth]{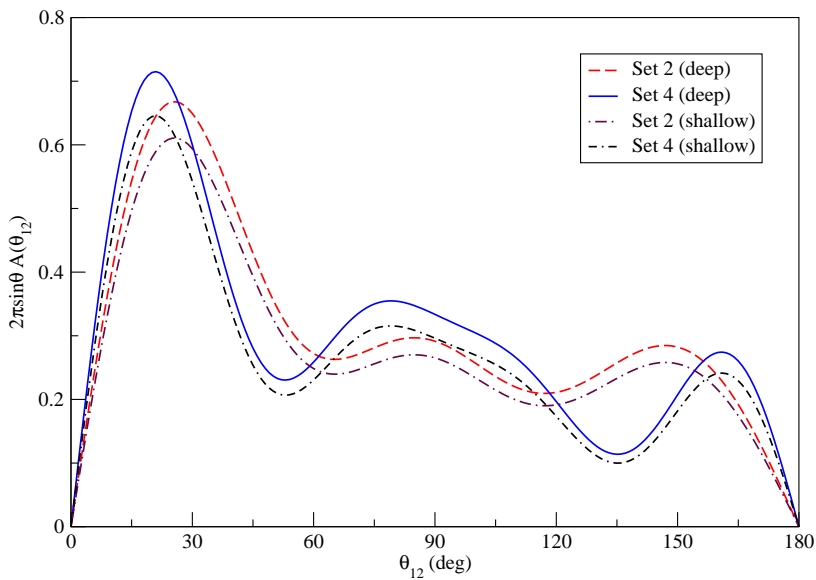}}
    \caption{(Color online) Angular density for $S=0$ weighted with a factor $2\pi\sin\theta_{12}$
for the ground state of the $^{22}$C as a function of the opening angle between the valence neutrons $\theta_{12}$ 
      for the shallow and deep case with the ground-state energy $-0.140$\,MeV and $-0.500$\,MeV respectively.
      Sets~$2$ and $4$ are employed.}
    \label{AD}
\end{figure}

The two particle density of $^{22}$C as a function of two radial coordinates, $r_1$ and $r_2$, 
for valence neutrons, and their opening angle, $\theta_{12}$ in the $LS$-coupling scheme is given by
\begin{equation}
\rho(r_1, r_2, \theta_{12})=\rho^{S=0}(r_1, r_2, \theta_{12})+\rho^{S=1}(r_1, r_2, \theta_{12})
\end{equation}
The explicit expression for $S=0$ component is given by \cite{BERT91}
\begin{align}
\rho^{S=0}(r_1, r_2, \theta_{12}=&\frac{1}{8\pi}\sum_{L}\sum_{\ell,j}\sum_{\ell',j'}\frac{\hat{\ell}\hat{\ell'}\hat{L}}{\sqrt{4\pi}} 
\begin{pmatrix}
  \ell & \ell' & L \\
  0 & 0 & 0 
\end{pmatrix}^2 \nonumber \\ 
&\times(-1)^{\ell+\ell'}\sqrt{\frac{2j+1}{2\ell+1}}\sqrt{\frac{2j'+1}{2\ell'+1}}\nonumber \\ 
&\times \psi_{\ell j}(r_1,r_2) \psi_{\ell'j'}(r_1,r_2) Y_{L0}(\theta_{12})
\end{align}
where $\hat{\ell}=\sqrt{2l+1}$ and $ \psi_{\ell j}(r_1,r_2)$ is the radial part of the two-particle wave function which is 
determined from Eq. (\ref{tpwfn}) by making use of Eqs. ($5$) and ($6$) of \cite{JS16}.
 The angular density for $S=0$ is defined by \cite{HAG05}
 \begin{equation}
  A(\theta_{12})=4\pi\int_{0}^{\infty}r_1^2 dr_1 \int_{0}^{\infty}r_2^2 dr_2\,\rho^{S=0}(r_1, r_2, \theta_{12}).
  \label{ADD}
 \end{equation}
   
Figures \ref{DenS} and \ref{DenD} shows the two-particle density plotted as a function of the radius $r_1=r_2=r$ and their opening
angle $\theta_{12}$, with a
weight factor of $4\pi r^2\cdot2\pi r^2$sin$\theta_{12}$ for Sets~$2$ and $4$ for the shallow and deep case, respectively. 
We also calculate the two-particle densities for Sets~$1$ and $3$, and find them to be consistent with the previous study 
\cite{HOR06}.
One can see in Figs. \ref{DenS} and \ref{DenD} that the two-particle density is well concentrated in 
the smaller  $\theta_{12}$, which is the clear indication of 
the di-neutron correlation. The distribution at smaller and larger $\theta_{12}$ are 
referred to as \textquotedblleft di-neutron\textquotedblright 
and \textquotedblleft cigar-like\textquotedblright configurations, respectively. 
The di-neutron structure component for both Sets~$2$ and $4$ has a relatively higher density, 
in comparison to the small cigar-like component. 
Figure~\ref{AD} shows the three peak decomposition of the angular density of Eq. ~(\ref{ADD}) as a function of $\theta_{12}$.
The origin of the central peak around $\theta_{12}=90\degree$ 
(\textquotedblleft boomerang\textquotedblright configuration \cite{HOR06A}) 
is from the configuration mixing with $d$- and $f$-waves coupled to even-$L\geq 2$ for the Sets~$2$ and $4$.
One can see the magnitude of the boomerang component is larger for the shallow and deep binding energy cases 
for Set~$4$ in comparison to Set~$2$, this effect can also be seen from Figs. \ref{DenS} and \ref{DenD}. 
The extension of the di-neutron structure is different for both sets and is attributed to the 
large degree of mixing of different
$l$-components for the Set~$4$ (see from Tables~\ref{CMC22a} and \ref{CMC22b}). 
We will explore this difference in the nuclear responses in the following sections. 

\section{Electric-dipole strength distribution}\label{DIP}

\begin{figure}[ht!]
  \centering
    \includegraphics[width=0.8\textwidth]{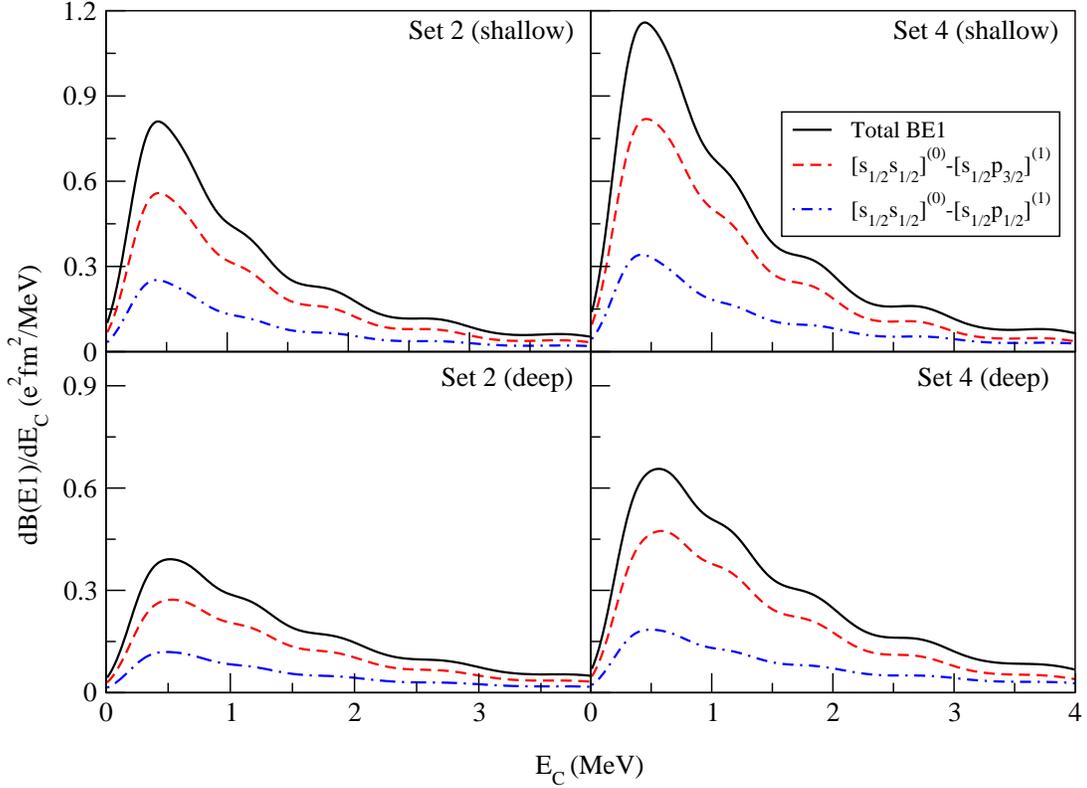}
    \caption{
      (Color online) Total $E1$ transition strength distribution 
      from the ground state $0^+$ to the final state $1^-$ for $^{22}$C along with the dominant transitions for Sets~$2$ and $4$
      for the shallow and deep cases with the ground-state energies of $-0.140$\,MeV and $-0.500$\,MeV, respectively.}
    \label{Dip2}
  \end{figure}
The detailed mathematical set up for calculations of the electric-multipole transitions to continuum 
is explained in Sec. $6$ of \cite{JS16}.
There is only a number of limited theoretical studies, 
that have been focused on calculations of the electric-dipole ($E1$) strength function 
\cite{PIN16,ERSH18}. In order to compare our results with these predictions, we also perform 
a set of calculations for the $E1$ response from the ground state to all components of $J^\pi=1^-$ state. 
The $E1$ excitations describes the transitions from the ground state to the continuum 
states by $E1$ operator \cite{JS16}.
After constructing a basis for sufficiently dense continuum, made up of eight 
components, namely $[s_{1/2}\times p_{1/2}]^{(1)}$, 
$[s_{1/2}\times p_{3/2}]^{(1)}$, $[p_{1/2}\times d_{3/2}]^{(1)}$, $[p_{3/2}\times d_{3/2}]^{(1)}$, 
$[p_{3/2}\times d_{5/2}]^{(1)}$, $[d_{3/2}\times f_{5/2}]^{(1)}$, $[d_{5/2}\times f_{5/2}]^{(1)}$ and 
$[d_{5/2}\times f_{7/2}]^{(1)}$, we diagonalize the Hamiltonian and obtain eigenvalues. 
We calculate the $E1$ response for Sets~$2$ and $4$  
and use the same value of the pairing strength that is used for the ground state.
There is total of $16$ possible different transitions from the 
initial $0^+$ ground state to the final $1^-$ state of $^{22}$C.
We investigate the detailed structure of the $E1$ strength distribution to study the role of different configurations. 
However, our model does not include the recoil correction at the moment. 
This was discussed in Ref.~\cite{ESBN97}, where it was concluded that the no-recoil 
approximation works quite well for several observables.
Figure~\ref{Dip2} shows the total $E1$ transition strength of $^{22}$C and 
its contribution of dominant transitions for the two different ${\rm core}$-$n$ potential 
Sets~$2$ and $4$, for the shallow and deep binding energies. 
Our calculations satisfy the cluster sum rule \cite{BERT92}. The total integrated strengths with the range of
$0<E_C<4$\,MeV accounts for $89.2\%$ ($93.9\%$) and $88.8\%$ ($95.5\%$) of the sum rule value 
for Set~$2$ and Set~$4$ for the shallow (deep) binding energy respectively. 
Our calculations give strength distributions at discrete values of energy to which we apply a Gaussian smoothing 
procedure that does not alter the total integrated strength.
As a result of the smoothing procedure, the curves in Fig.~\ref{Dip2} show a few minor wiggles, that are not to be attributed to resonances, 
but must be considered as an artifact. 
The shape and strength of our $E1$ response function are consistent with the previous calculations \cite{ERSH12,PIN16,ERSH18}.
It is clear, though, that there is a concentration of strength at low energies and possible maximum close to $\sim0.4$ and $\sim0.6$\,MeV 
for the shallow and deeper case respectively.
We find in these calculations that the transitions $[s_{1/2}\times s_{1/2}]^{(0)}$ $\rightarrow$ $[s_{1/2}\times p_{3/2}]^{(1)}$ 
and $[s_{1/2}\times s_{1/2}]^{(0)}$ $\rightarrow$ $[s_{1/2}\times p_{1/2}]^{(1)}$ 
play the dominant role in the total $E1$ transition strength and are shown in the Fig.~\ref{Dip2} for both sets, 
whereas all the remaining $14$ transitions are less significant. 
The role of the core excitation has been discussed \cite{INAK14} in context to the momentum distribution of 
$^{20}$C fragment of $^{22}$C breakup, which is beyond the scope of the present study.
Our findings show the negligible difference in $E1$ response for two different ${\rm core}+n$ potentials, 
for the same binding energy. The negligible difference for the peak position 
with different ${\rm core}+n$ potentials is due to the fact that
the two sets give the same binding energies. The shifting of the peak of the $E1$ strength distribution 
with different binding energies was discussed in \cite{PIN16,ERSH18}. 
We also find the slight shift of the peak position to the higher energy side for the deep case.
Also, for the deep case due to more mixing of $d$- and $f$-waves, the $E1$ strength distribution 
is wider than that corresponding to the shallow case. 
In order to reach final conclusion for the $E1$ response, precise measurement of $^{22}$C  binding energy is 
highly needed of the moment.

\section{Monopole strength distribution}\label{MON}
\begin{figure}[ht!]
  \begin{center}
    \includegraphics[width=0.8\linewidth]{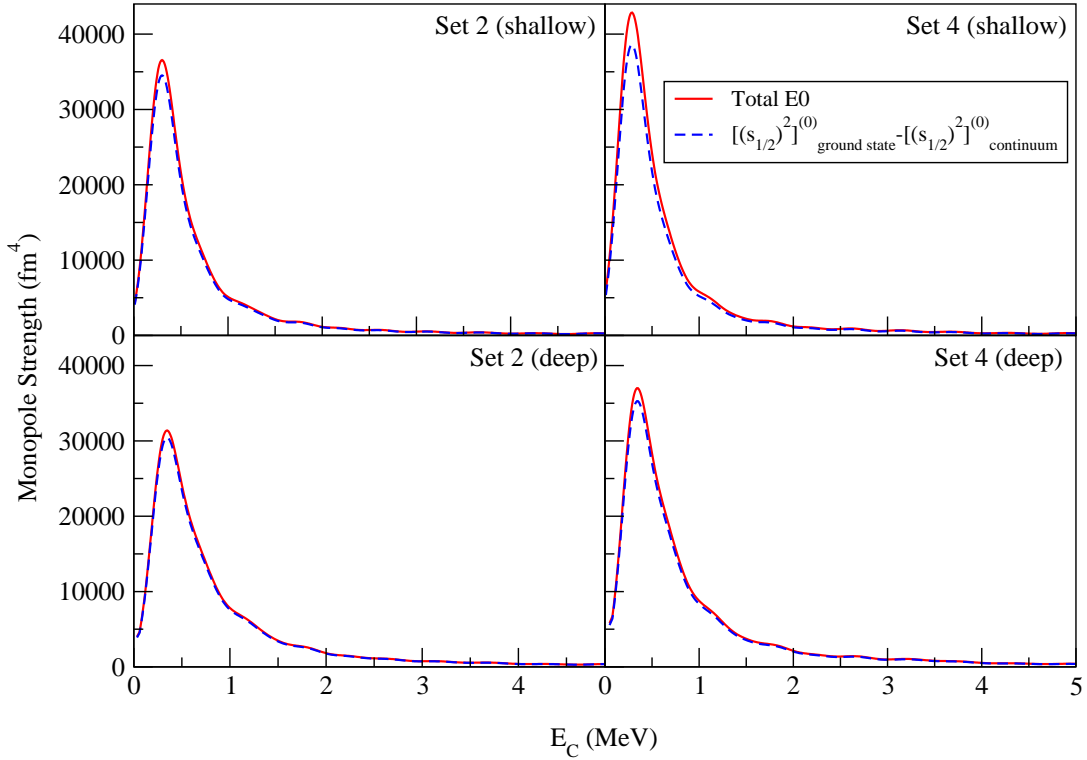}
    \caption{
      (Color online) Total monopole transition strength distribution and the dominant transition
      from $0^+$ (ground state) to the $0^+$ (continuum) for $^{22}$C for Sets~$2$ and $4$ 
      for the shallow and deep cases with the ground-state energies of $-0.140$\,MeV and $-0.500$\,MeV, respectively.}
    \label{Mon3}
  \end{center}
\end{figure}
Among the excited states in weakly-bound nuclei, the
excitation modes which correspond to the collective motion between halo neutrons  
and core are particularly intriguing. This section is devoted to study 
the low-energy monopole excitations in $^{22}$C.
The monopole excitations describe the transitions from the $J^\pi=0^+$ ground state 
to the $J^\pi=0^+$ continuum states with
monopole operator defined by $\sum_{k=1}^2r_{k}^2$.
These calculations have also led us to study the role of various configurations in the total monopole strength.
After constructing a basis made up of seven components, namely $(s_{1/2})^2$, $(p_{1/2})^2$, $(p_{3/2})^2$, $(d_{3/2})^2$ 
$(d_{5/2})^2$, $(f_{5/2})^2$ and $(f_{7/2})^2$, we diagonalize the Hamiltonian for $J^\pi=0^+$. 
Only one state is bound and all the remaining ones are unbound. 
There are seven possible transitions from the ground-state components to continuum $0^+$ states ones.
Figure~\ref{Mon3} shows the total monopole transition strength of $^{22}$C and the contribution 
of the dominant transitions for Sets~$2$ and $4$, for the shallow and deep binding energy. 
Out of the possible seven transitions, $[(s_{1/2})^2]^{(0)}$(ground state)$\rightarrow$ $[(s_{1/2})^2]^{(0)}$(continuum) 
transition is the dominant one, 
whereas the transition $[(f_{5/2})^2]^{(0)}$(ground state)$\rightarrow$ $[(f_{5/2})^2]^{(0)}$(continuum) is the 
least significant in the total monopole transition strength.
From Fig.~\ref{Mon3}, one can see there is slightly larger difference between the total monopole strength and the dominant 
transition strength for Set~$4$ than Set~$2$. This is due to the fact that for Set~$4$, we have slightly larger 
contribution from the two other less significant transitions, i.e., $[(d_{3/2})^2]^{(0)}$(ground state)$\rightarrow$ $[(d_{3/2})^2]^{(0)}$(continuum) and
$[(f_{7/2})^2]^{(0)}$(ground state)$\rightarrow$ $[(f_{7/2})^2]^{(0)}$(continuum). 
This fact can also be seen from large percentage of the $d_{3/2}$-wave $f_{7/2}$-wave for potential Set~$4$ (see Table ~\ref{CMC22a}), for the shallow case.
The peak position for the total monopole strength, for these two different sets is almost same as they correspond to the same binding energy. 
We find the slight shift in the peak position to higher energy for the deep case. 
Also, the magnitude of the peak strength is larger for Set~$4$ as compared to Set~$2$, this can be due to the fact that
Set~$4$ gives larger matter radius than that of Set~$2$. Again, one needs accurate binding energy to reach the 
final conclusion for the low-lying continuum spectrum of $^{22}$C.
\section{Summary and Conclusions}\label{CON}
In summary, the sensitivity of the core-neutron potential has been explored with the mixing of different $l$-wave components
in the ground state of $^{22}$C by using a three-body structure model \cite{FORTU14,JS16,JS16b,JS16c}. 
The ground-state properties have been calculated and found to be consistent with 
other theoretical predictions. Also, the electric-dipole and monopole responses of $^{22}$C have been investigated and the role of the various possible 
transitions has been studied in each case.

Our results strongly depend on the choice of the ${\rm core}+n$ potential and 
the binding energy of the ${\rm core}+n+n$ system.
Four different potential sets have been examined and the emergence of the $f$-wave resonance
has been discussed. Sets~$1$ and $3$ are examined to check the consistency of our calculations with the previous studies, 
whereas Sets~$2$ and $4$ predict the different position of the $f$-wave resonance state. 
In order to explore configuration mixing we have studied the electric-dipole and monopole responses. We find that there is negligible difference 
in the response functions for the same binding energy. 

In addition, our results for Set~$4$ predict the $s$-wave dominance for the lower binding energy ($-0.140$\,MeV) and 
the $f$-wave dominance for the deeper binding energy ($-0.500$\,MeV). 
However, we found that it is difficult to propose the best ${\rm core}+n$ potential unless 
we have more precise experimental data for the low-lying spectrum of $^{21}$C as well as for the binding energy of $^{22}$C.
Thus, our results encourage the precise experimental measurements for these observables. 
In the present calculations we assumed the inert $^{20}$C core, it would be interesting to extend our model 
by considering the core excitations as a future direction. We expect that our efforts might be of help to unravel 
the structure of low-lying continuum spectrum of $^{22}$C.

\section*{Acknowledgements}
We would like to thank  P. Descouvemont and J. A. Lay for useful discussions. 
J. Singh gratefully acknowledged the financial support from Nuclear Reaction Data Centre (JCPRG), Faculty of science, Hokkaido University, Sapporo. 
This work was in part 
supported by JSPS KAKENHI Grant Numbers 18K03635, 18H04569 and 19H05140, and the collaborative research program 2019,
information initiative center, Hokkaido University.

\section*{Appendix - Convergence with model parameters}
\label{CON1}

\begin{figure}[th]
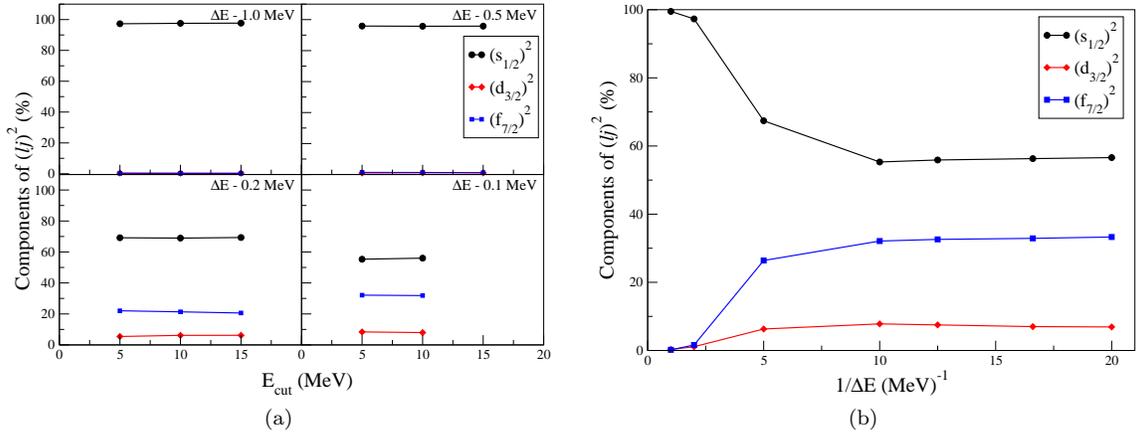

\vspace{0.2cm}
  \begin{center}
   \subfigure[] {\epsfig{file=Fig6a.eps, scale=0.30}}
   \hspace*{0.3cm}
    \subfigure[] {\epsfig{file=Fig6b.eps, scale=0.30}}
    \caption{
      (Color online) (a)Convergence with continuum energy cut $E_{\rm\,cut}$ of the model 
      space and (b)Convergence with energy spacing $\Delta{E}$ of the model space.}
    \label{22}
  \end{center}
\end{figure}
We calculate the continuum single-particle wave functions for different $E_{\rm\,cut}$'s. 
After fixing the convergence with $E_{\rm\,cut}$, the continuum single-particle wave functions, 
using the mid-point method for $E_{\rm\,cut}=5$\,MeV, with $\Delta{E}=1.0$, $0.5$, $0.2$, $0.1$, $0.08$, $0.06$ and $0.05$\,MeV,
the two-particle states are formed and the matrix elements of the pairing interaction are calculated.
The convergence of the results is discussed on two different levels:
\begin{enumerate}
\item  Convergence with the continuum energy cut ($E_{\rm\,cut}$) of the model space\\
We check the convergence of our results (dominant contribution in the ground state of $^{22}$C) with different 
$E_{\rm\,cut}$'s. In the Fig.~\ref{22}(a), we show the variation of 
three dominant contributions ($s_{1/2}$, $d_{3/2}$ and $f_{7/2}$) in the ground state of $^{22}$C with different $E_{\rm\,cut}$'s for different 
values of the energy spacing ($\Delta{E}$). 
One can clearly see that we achieve the convergence with different $E_{\rm\,cut}$'s but still there is significant 
variation in the percentage contribution with different $\Delta{E}$s. In the fourth quadrant of the Fig.~\ref{22}(a), 
due to huge computational costs, we restrict ourselves to only $10$ MeV. It is convincing that one can 
choose fairly well the $E_{\rm\,cut}=5$\,MeV for this study and we will do all the calculations with $E_{\rm\,cut}=5$\,MeV.
\item Convergence with the energy spacing ($\Delta{E}$) of the model space\\
As seen from the Fig.~\ref{22}(b), 
we study the convergence of our results 
(dominant contribution in the ground state of $^{22}$C) with inverse of $\Delta{E}$ for $E_{\rm\,cut}=5$\,MeV. 
It is clear from the Fig.~\ref{22}(b), the curves start getting flat from  $\Delta{E}=0.1$\,MeV. 
Therefore, we adopt $\Delta{E}=0.1$\,MeV ($1/\Delta{E}=10$\,MeV) for the present calculations.   
\end{enumerate}

\end{document}